\documentclass[sn-basic]{sn-jnl}% Basic Springer Nature Reference 
\theoremstyle{thmstyleone}%

\theoremstyle{thmstyletwo}

\theoremstyle{thmstylethree}

\usepackage[utf8]{inputenc}
\usepackage{physics}

\begin{document}
\pagestyle{plain}

\title{Investigating internal migration with network analysis and latent space representations: An application to Turkey}

\author*{\fnm{Furkan} \sur{Gürsoy}}\email{furkan.gursoy@boun.edu.tr}

\author{\fnm{Bertan} \sur{Badur}}\email{badur@boun.edu.tr}

\affil{\orgdiv{Dept. of Management Information Systems}, \orgname{Boğaziçi University}, \orgaddress{\street{Bebek}, \city{Istanbul}, \postcode{34342}, \country{Turkey}}}

\abstract{    Human migration patterns influence the redistribution of population characteristics over the geography and since such distributions are closely related to social and economic outcomes, investigating the structure and dynamics of internal migration plays a crucial role in understanding and designing policies for such systems. We provide an in-depth investigation into the structure and dynamics of the internal migration in Turkey from 2008 to 2020. We identify a set of classical migration laws and examine them via various methods for signed network analysis, ego network analysis, representation learning, temporal stability analysis, community detection, and network visualization.  The findings show that, in line with the classical migration laws, most migration links are geographically bounded with several exceptions involving cities with large economic activity, major migration flows are countered with migration flows in the opposite direction, there are well-defined migration routes, and the migration system is generally stable over the investigated period. Apart from these general results, we also provide unique and specific insights into Turkey. Overall, the novel toolset we employ for the first time in the literature allows the investigation of selected migration laws from a complex networks perspective and sheds light on future migration research on different geographies.}

\keywords{internal migration, Turkey, network analysis, representation learning, community detection}

\maketitle

\section{Introduction}

Human migration takes place for a large variety of reasons including those that are economic, social, educational, cultural, religious, safety-related, and so on. Most generally, migration can be defined as a relatively permanent move from one migration-defining area to another \citep{pieter1997}. A distinction between two types of migration can be made based on whether they are international or between settlements within the same country. World Bank estimates that 272 million people are international migrants in 2020, which corresponds to 3.5\% of the world population with an accelerating increase from 2.3\% in 1970 and from 2.8\% in 2000 \citep{wb2019}. Still, the vast majority of people live in the countries they were born. On the other hand, internal migrants are estimated at 740 million people in 2009 \citep{undp2009}, corresponding to more than 10\% of the world population.

\pagestyle{plain}

The research literature on migration is concerned with questions on \textit{who}, \textit{why} (e.g., the determinants), \textit{where}, and \textit{when} of migration and its consequences \citep{greenwood1997}. Another distinction of the literature can be made based on whether the migration is studied at a personal (i.e., microscale) or aggregate level (i.e., macroscale) \citep{danchev2020migration}. The migration inflows and outflows affect the (re)distribution of population and its characteristics over the geography. As such distributions are closely related to social and economic outcomes, investigating the structure and dynamics of migration are of (or should be of) utmost importance to demographic and developmental policymakers.

Lately, the application of network analysis tools and concepts to migration data has become more popular as they can viably analyze \textit{where} of migration with an overview of the whole system while accounting for the interdependencies in its structure. Methods from the representation learning literature have not been employed in the migration literature so far.

In this study, we focus on internal migration at an aggregate level focusing mainly on the \textit{where} and \textit{when} question with a temporal network analysis approach and a representation learning-based latent space model. We extend the frequently employed network analysis methods by employing a filtering method that extracts not only positive migration ties but the negative ties (i.e., a signed network) and by employing methods from the machine learning literature to learn and analyze the latent low dimensional representation of the migration system. We provide our analysis on the dynamic internal migration flows in Turkey from 2008 to 2020 although the same methods apply to all such internal migration systems. To the best of our knowledge, this study is the first application of signed networks and representation learning methods to migration networks, and the first extensive network analysis study for the internal migration system in Turkey,

In the seminal works of \cite{ravenstein1885, ravenstein1889} and \cite{lee1966}, several laws are proposed as facts and governing rules of migration. Below, we extract and discuss some of those laws that can be addressed with the data that we have.

\begin{itemize}

    \item Migration tends to show an increasing trend in terms of volume and rate unless stringent counter-policies are implemented. In contrast, the neoclassical economic view suggests that migration tends to neutralize the economic pull and push factors; and as such differences diminish, the migration tends to decrease over time \citep{neoclassic}.
    
    \item Migration takes place in well-defined routes from specific source locations to specific destinations. Popular examples include large numbers of Southern Italian immigrants in South Philadelphia \citep{southernitaly}, Vietnamese immigrants in Prague \citep{vietnam}, Japanese immigrants in Brazil \citep{japanese}, and the large number of immigrants from particular small towns of Turkey such as from Kulu, Emirdag, and Unlupinar to Sweden, Belgium, and London, respectively \citep{kulu, emirdag, unlupinar}. Such highly specific routes gain strength due to the knowledge and support reaching back from the destination due to personal social networks and other factors alike. In addition to the specific routes, this law also states that some specific destinations are highly attractive, therefore, attract much larger migration streams, which indicates a power law in migration flow volumes.

    \item For every major migration flow, another stream develops in the opposite direction. Such counter-streams exist due to contact created between the origin and destination as well as return migration. \cite{rogers1983} enumerate eight types of reasons for return migration based on events related to origin or destination, at the aggregate or individual level, and related or unrelated to the original motive for migration.
    
    \item Most migrants only move smaller distances while those who migrate longer distances usually do so towards centers of large economic activity.
\end{itemize}

The remainder of this study is structured as follows. In Section \ref{sec:lr}, we provide an overview of migration research with a specific focus on network analysis approaches at an aggregate level for internal migration systems. In Section \ref{sec:mm}, we describe our dataset, introduce the mathematical notation we employ, and explain network analysis, information filtering, representation learning, and community detection methods that we utilize in this study. In Section \ref{sec:res}, we present the results of our analysis in line with but not limited to the above-mentioned laws and discuss our findings. We provide brief conclusions and final remarks in Section \ref{sec:con}.

\section{Related work}\label{sec:lr}

The earliest study on internal migration that we have identified is the seminal study of \cite{ravenstein1885} that investigates the migration patterns within the United Kingdom and provides extensive hand-drawn cartographies detailing the migration patterns. Early works of \cite{slater1974college, slater1975, slater1975identification, bistochastic, slater1976spain} may be considered the first studies in the migration literature that employs an explicit networks perspective. However, a wide range of network analysis tools are relatively recent and have not found much use in the migration literature. \cite{BILECEN20181} remark that although network characteristics of migration are widely acknowledged, network analysis methods have not received the deserved attention. \cite{caudillo2014} also highlight the lack of a network analysis approach in migration apart from few studies.

Although internal migration data is available at a more fine-grained and reliable level, it has not attracted as much attention as international migration \citep{carvalho2020}, particularly in geographies outside United States \citep{Pitoski2021}. An early work to employ network analysis perspective studies the migration patterns at the state level in the US \citep{gunter2005}. They highlight the problem of having very dense networks, and accordingly, the need for filtering methods. 
The study identifies the hierarchical community structure and finds that the communities are geographically bounded. \cite{slater2008hubs} investigates the hub and cluster structure in the US at the county level for two time periods. The study also identifies the hierarchical community structure and finds that cosmopolitan areas serve as hubs, have longer-range links, and tend to connect to other geographically bounded clusters at upper levels rather than forming immediate close-knit clusters.  \cite{goldade2018} compare the internal migration in the US at the county level during the housing boom (2004-2007) and the following recession (2008-2011) in terms of their network structure, geographical bounds and political affiliation of communities, and its dynamic behavior. The study finds that high degree nodes do not have high clustering whereas low degree nodes generally show high clustering, migration dynamics are largely stable over time, and the communities are geographically bounded. \cite{Charyyev2019} also provide a similar analysis from 2000 to 2015 at the county level with comparisons of periods of economic prosperity and recession. They investigate the dynamic behavior by monitoring the existence of links over time and find considerable instability. However, it might be due to the very granular view of the network as they also found that the migration network maintained its characteristics over the investigated period, in constrast to their instability result.

Outside the US, the studies on internal migration with a network analysis approach are usually more recent. \cite{caudillo2014} compare the migration patterns of highly qualified and general population between 2005 and 2010 in Mexico at the level of metropolitan areas and provide an online visualization tool for further analyses. \cite{Pitoski2021} study the internal migration in Croatia and find that both degrees and strengths follow a power-law distribution. They also find that migration flows show a reciprocatory behavior and geographically bounded communities arise. \cite{chen2021} provides a visual network analysis of internal migration within England and Wales in 2019 and find that people tend to move within groups of geographically close regions. They further indicate that such visualizations can support informed planning and decision-making on policies regarding public services, economic and social development, and alike. \cite{carvalho2020} study dynamics of migration networks from 1980 to 2010 in Brazil and find high reciprocity and hierarchical network topology. In line with other cities, they also find that long-range links generally belong to larger cities, new migration flows tend to be between pairs with previously established migration flows.

The earliest study that we identified which employs a network analysis approach to the internal migration in Turkey is of \cite{slater1975identification}. The study identifies migration flows between then 67 cities by looking at the discrepancies in the cities people were born and live.  The study finds the following clusters of cities: Amasya, Tokat, Çorum, Sivas;  Kayseri, Kırşehir, Nevşehir, Yozgat; Kocaeli, Sakarya, Bolu; Bilecik, Bursa, Eskişehir;  Aydın, Muğla, Denizli; Kütahya, Uşak; Antalya, Burdur, Afyonkarahisar, Isparta; Edirne, Kırklareli, Tekirdağ; Hakkari, Van; Ağrı, Bitlis, Kars, Muş; Diyarbakır, Mardin, Siirt; Bingöl, Tunceli, Elazığ; Artvin, Rize; Adana, Adıyaman, Gaziantep, Hatay, Mersin, Malatya, Kahramanmaraş, Şanlıurfa; Giresun, Kastamaonu, Ordu, Samsun, Sinop, Zonguldak, Çankırı; and Trabzon, Gümüşhane. It is noted that Istanbul and Ankara are not strongly tied to any community. Apart from this, to the best of our knowledge,  the work of \cite{yakar2017} is the only study that takes a network analysis approach at an aggregate level to study internal migration in Turkey. However, their data is based on the birth place, hence providing a snapshot rather than dynamic flows, as is the case with the study of \cite{slater1975identification}. They investigate statistics like degree and centrality measures under different levels of global thresholds for link weight. Their findings show that the links towards Istanbul are not affected much by distance whereas other links carry a more regional/local characteristic.

Overall, our review of the literature finds that the studies that employ a network analysis approach to investigate internal migration are limited, usually very recent, and particularly focused on the US. Findings indicate that geographically bounded communities form a hierarchical structure with anomalous links involving areas with large social and economic activity. Migration flows tend to be reciprocated and well-defined streams exist. The migration networks tend to be stable over time even during times of economic prosperity and recession. We also note that even very recent studies employ global thresholds than more appropriate network sparsification methods, methods from the machine learning literature have not yet found a use, and network analysis studies on internal migration at the aggregate level are not only rare in Turkey but also in other countries outside the US.

\section{Material and methods}\label{sec:mm}

\subsection{Dataset}

The dataset in this study is sourced from publicly available migration data published by the \cite{tuik2021}. The migration figures are calculated based on change of official residences in the Adress Based Population Registration System (ABPRS) which typically indicates a relatively permanent relocation. Specifically, a migration is said to have occurred if the city of official residence is changed as compared to the previous year. As information in ABPRS is utilized extensively in most tasks of the citizens including voting, utility services and billing, and so on, we consider the dataset to be sufficiently complete.

The data is available at the level of cities at the most granular level. There are 81 cities in Turkey with the minimum, median, mean, and maximum populations of respectively about eighty thousand, half a million, one million, and sixteen million out of a total population of 83.6 million as of 2020. At more general levels than the city level, there are geographic regions and divisions in addition to more specific groupings that are created exclusively for census/statistical purposes. Being a strictly unitary state, there are no administrative divisions above the city level. Below the city level, there are districts and neighborhoods/villages, for which migration flows are not available. We further believe that investigating the internal migration at the city level is appropriate as districts and neighborhoods are not well-separated in terms of economic and even social activities. As migration usually requires a minimum distance of movement, we believe it is an appropriate choice of migration-defining area.

ABPRS became effective in 2007, therefore, the migration data is available from 2008 up until 2020. 81 cities remain the same throughout the investigated period. ABPRS covers all Turkish citizens and foreign nationals residing in the country. However, the internal migration data excludes foreign nationals as well as those who are relocated due to mandatory military service or serving their prison sentences, and citizens who spend less than six months in Turkey over a year.

We model the migration in-flows and out-flows between cities as a directed and weighted network. For each year, we generate a separate network. Thus, in each network, a link denotes a migration flow from an origin city (i.e., a source node) to a destination city (i.e., a target node) for that year. Its weight corresponds to the number of people that migrate from the source to the target node. Consequently, we obtain 13 networks and each network has 81 nodes and the migration links between those cities.

\subsection{Notation}
Formally, we define a directed and weighted network with $G(V, E, W)$ where $V$ is the set of nodes with its general members $i, j, \dots$ and $E$ is the set of directed edges. $\abs{V} = n$ is the number of nodes and $\abs{E} = m$ is the number of edges. $W$ is the weighted adjacency matrix where $W_{ij}$ is the number of migrants from $i$ to $j$ and $W_{ii} = 0$. In-degree and out-degree of a node correspond to the number of cities that the city receives migration from and sends migration to, respectively. In-strength and out-strength of a node correspond to the number of migrants who arrived at the city and the number of migrants who left the city, respectively, i.e., the total in- and out-flow of the city.

In general, we denote scalars with lowercase letters (e.g., $x$), matrices with uppercase letters (e.g., $X$), an element of a matrix by two subscripts respectively corresponding to its row and column indices (e.g., $X_{ij}$), and row vectors of a matrix with lowercase letters followed by a single subscript (e.g., $x_i$). $\|x_i\|$ denotes the norm of $x_i$.

\subsection{Information filtering}

Density is defined as the proportion of observed links to the possible links. Here, as we deal with a directed network without self-loops, the density can be calculated as shown in Equation \ref{eq:density}. A density value of $1$ indicates a complete network where all possible links are observed. Too dense networks are undesired for various interrelated reasons. First, some of these links may not be significant or important hence may distort our understanding of them, e.g., when only a few people migrate from a city to another, can we faithfully consider it as a link between the two cities? Second, many network analysis methods desire or require sparsity for accurate or effective analysis. Third, insightful and apprehensible visualizations cannot be devised when there are too many and possibly insignificant and uninformative links. To this end, the literature, including those that study migration networks, utilize various filtering methods.

\begin{equation}\label{eq:density}
\frac{m}{n(n-1)}
\end{equation}

A filtering method is to establish a global threshold and eliminate links whose weights are below this threshold. This has two major problems. First, the choice of the threshold may be arbitrary and not justified by a theory or empirical evidence. Second, such global threshold methods are inappropriate for multiscale networks. The population of cities is distributed unevenly. Some cities are much more attractive and other cities are much more unattractive. As a result, few cities have much larger in- and/or out-flows and many cities have much smaller in- and out-flows in terms of absolute numbers. When a global threshold is applied, some informative links that connect, for instance, less populated cities may be eliminated while some uninformative links that connect well-populated cities may be retained. Despite such limitations, the global threshold method is applied by many studies in the literature including those by \cite{chen2021, carvalho2020, Pitoski2021, yakar2017}.

The multiscale nature (i.e., the heterogeneous scale of node strengths and edge weights) of migration networks is a widely observed phenomenon. Many empirical networks in various domains exhibit a multiscale nature in their edge weights and node strengths. Hence, the information filtering methods to extract network backbones are widely studied in the literature. Accordingly, \cite{gunter2005, slater2008hubs, Charyyev2019} use more sophisticated thresholding methods, e.g., bistochastic filter \citep{bistochastic} or disparity filter \citep{disparity}.

The null model of the disparity filter assumes that the normalized link weights of a node follow a uniform distribution. Hence, it is possible to extract statistically significant links by comparing the observed weights to the null model. The bistochastic filter also employs a normalization scheme but at the global level, and then identifies the links with the largest normalized values. The similarities and differences between the two methods are discussed by \cite{Slater2009} and \cite{SerranoResponse}. A common point of these methods and most other methods in the literature is the fact that only the links that are significant in the positive direction are extracted. However, nodes may have negative links between them which indicate dissimilarity, animosity, and alike. In addition to the positive links, migration networks may have underlying negative links, e.g., a usually low number of migrants between two cities with relatively large numbers of inflows and outflows.

Extracting negative links from networks with positive edge weights requires an assumption that links with too small edge weights or lack of links may actually correspond to negative links. Such an assumption does not hold in many networks but is acceptable in intrinsically dense networks. Intrinsically dense networks refer to those networks where "all nodes are aware of all other nodes and can interact with them without obvious natural limits" \citep{gursoy2020extracting}. As we are investigating a migration system that is empirically very dense (almost complete, in fact) and within a single country with free movement between its cities without limits like a different language, we may safely assume that it is an intrinsically dense network.

After establishing the intrinsically dense nature of the studied network, we use a recent filtering method \citep{gursoy2020extracting} that extracts not only the positive links but also negative links. The method works by building a null model that controls for the in-strength and out-strength of nodes, estimates the expected values of link weights under this model, and identifies links whose weights are sufficiently distant from their null expectations as positive or negative links. The method can extract the signed backbone at a desired level of sparsity based on statistical significance of the links (via its significance filter) as well as based on the intensity of the links (via its vigor filter). The intensity (vigor) of the links is a lift-based measure that ranges within $[-1,1]$ and may also be used as signed edge weights.

Signed networks, to an extent, usually manifest a specific type of reciprocity where positive links are reciprocated with positive links, and negative links with negative links \citep{gursoy2020extracting}. Conflicting links where a positive link is reciprocated with negative links are usually rare. Reciprocity in directed networks can be defined as the proportion of directed links that are reciprocated. For signed networks, we can extend the definition with a further restriction that requires reciprocated links to be of the same sign, i.e., ratio of reciprocated links that are of the same sign over all links. Similarly, one can calculate the ratio of conflicting links over all links.

Most empirical signed backbones also exhibit structural balance (SB) to some extent. Structural balance can be intuitively summarized with the phrases "friend of a friend is a friend" and "enemy of a friend is an enemy". This phenomenon can be thought of in relation to the local clustering of nodes in binary networks where a node tends to be linked with its neighbors' neighbors. \cite{cartwright} formalized structural balance for an undirected triple of nodes to be balanced if edge signs between them are {+, +, +} or {+, -, -} and unbalanced if the edge signs are {+, +, -} or {-, -, -}. \cite{davis} suggested that an enemy of a friend is not necessarily an enemy and proposed weak structural balance (WSB) where the only unbalanced triple is the one with edge signs {+, +, -}. Consequently, SB and WSB of a signed network are defined as the proportion of balanced triples over all triples.

\subsection{Dynamic representation learning}

The information that networks present are usually encoded in adjacency matrices with $n$ rows and $n$ columns. Such representations may suffer from high dimensionality. To overcome such problems, the goal of representation learning (RL) is to learn low dimensional latent representations (i.e., embedding vectors) of nodes such that most of the information in the observed network is preserved in lower dimensionality. To this end, the similarity in the embedding space should approximate the similarity in the observed network. Alternatively, the representations in the embedding space can be seen as the underlying true data that generates the observed network. Formally, the problem is to estimate embedding matrix $Z$ with shape $(n\times d)$ where $n$ is the number of nodes, $d$ is the embedding size, and $z_i$ is the embedding vector of the node $i$ such that $f(i, j) \sim g(z_i, z_j)$ where $f$ is a similarity function in the observed network and $g$ is a similarity function in the latent space.

Human migration, as modeled by networks, is well-suited for RL since the low-dimensional representation of nodes (e.g., cities in this case) enables further analysis, for instance, via methods for traditional clustering algorithms. Furthermore, it may be more tractable to monitor the dynamics of the system in lower dimensions.  

We describe our representation learning method as follows. Given $G$, we first calculate the underlying signed backbone network $\hat{G} = (\hat{V}, \hat{W})$ by setting both significance filter and vigor filter to $0$, i.e., retaining all possible edges.  $\hat{W}$ is the weight matrix whose elements are in the range $[-1,1]$. We employ $\hat{W}$ as the similarity in the observed network where $\hat{W}_{ij}$ denotes the similarity of nodes $i$ and $j$, i.e., $f(i,j) = \hat{W}_{ij}$. The network is directed thus $W_{ij} \neq W_{ji}$, however, our optimization process will find embeddings that are optimized for an appropriate trade-off between the two as will be obvious shortly.

To calculate similarity in the latent space, there are multiple alternatives such as euclidean distance, dot product, and cosine of the angle between two embedding vectors. Cosine similarity is defined in Equation \ref{eq:cossim}. Since the range of similarity in the input data (i.e., edge weights in the signed backbone) is $[-1, 1]$, we choose cosine similarity whose natural range is also $[-1, 1]$, i.e., $g(z_i, z_j) = cossim(z_i, z_j)$. As the angle between two points in the latent space increases towards $180^{\circ}$, its value approaches $-1$. As the angle decreases towards $0^{\circ}$, its value approaches $1$. When they are orthogonal, its value is $0$. When $d = 2$, the embedding vectors are located on a circle in the latent space. When $d = 3$, the embedding vectors are located on the surface of a sphere. Similarly, when $d > 3$, the embedding space is the surface of a higher-order sphere.

\begin{equation}\label{eq:cossim}
    cossim(z_i, z_j) = \frac{z_i z_j}{\|z_i\| \cdot \|z_j\|}
\end{equation}

After we establish the similarity functions in the observed space and latent space, we define the loss function $\mathcal{L}$  as in Equation \ref{eq:loss}. We optimize $Z$ using stochastic gradient descent \citep{sgd1951}. We use stochastic gradient descent to find $Z$ which minimizes $\mathcal{L}$. After initiating $Z$ with random values, we iterate over each non-diagonal element of $W$ in random order, calculate respective gradients for $z_i$ and $z_j$, and update them using a learning rate $\alpha = 0.01$. Each iteration over all elements of $W$ is called an epoch. For each year's network, this process is repeated for 100 epochs.

\begin{equation}\label{eq:loss}
    \mathcal{L} = \sum{\abs{cossim(i, j) - W_{ij}}}
\end{equation}

More specifically, we need to find the gradient vector to update the embedding vector $z_i$ for each iteration over all instances of $W_{ji}$ or $W_{ij}$ where $j \in V-\{i\}$.

\begin{equation}\frac{\partial} {\partial{z_i}} \abs{W_{ij} - cossim(z_i, z_j)}\end{equation}

For simple presentation, we define the quantity $y$ as follows.

\begin{equation}y = W_{ij} - cossim(z_i, z_j)\end{equation}

Then, the problem becomes

\begin{equation}\frac{\partial}{\partial z_i}\abs{y} = \frac{y}{\abs{y}} \frac{\partial y}{\partial z_i}\end{equation}

where 

\begin{equation}\frac{\partial y}{\partial z_i} =  - \frac{\partial}{\partial z_i} cossim(z_i, z_j) 
  = - \frac{z_j}{\|z_i\| \cdot \|z_j\|} + \frac{z_i}{\|z_i\|^2} cossim(i,j)\end{equation}

We update $z_i$ as follows

\begin{equation}z_i \gets z_i - \alpha \frac{\partial}{\partial{z_i}}\abs{W_{ij} - cossim(z_i, z_j)}\end{equation}

and after each epoch, we ensure that the norm of $z_i$ is equal to $1$ via the update shown in Equation \ref{eq:upd}. Note that such normalization does not change the value of $cossim(z_i, z_j)$ function.

\begin{equation}\label{eq:upd}z_i \gets \frac{z_i}{\|z_i\|}\end{equation}

As implied by the loss function $\mathcal{L}$, the information in the latent space is stored in the relative positions of embedding vectors rather than their absolute positions. The optimization may converge to different but equally plausible solutions due to the random initialization of $Z$ or inherent stochasticity of the optimization procedure. For instance, any rotation and/or reflection of the found solution provides exactly the same information. It naturally follows that we cannot directly compare the embedding vectors between different time steps. There might be optimal rotations and/or reflection between different years' embeddings that preserve the cosine similarity while transforming the space. This phenomenon is known as procrustes problem \citep{procrustes} and discussed in-depth by \cite{gursoy2021alignment}. For each consecutive year, we find the optimal rotation matrix $R$ (as in Equation \ref{eq:rot}) and update the embedding matrix as $Z^{t+1} \gets Z^{t+1} R$ where $Z^t$ is the embedding matrix at year $t$. After this procedure, all $Z$ matrices are considered to be \textit{aligned} across all time steps, i.e., they are brought into the same latent space and are comparable.

\begin{equation}\label{eq:rot}
    R = argmin_{\Omega} \|Z^{t+1} {\Omega} - Z^t\|
\end{equation}

Stability reflects the changes between embeddings in two time steps that are not attributable to the above-described misalignment issues. Therefore, stability error \citep{gursoy2021alignment} reflects the extent of structural change in dynamic embeddings. Its range is $[0, 1]$ where $0$ indicates that the information is exactly the same between two embeddings and $1$ indicates the worst case. Therefore, lower values indicate a largely stable system.

\subsection{Community detection}

We identify community structures in the migration networks using two different clustering approaches: density-based and hierarchical. For the former, we employ Density Based Spatial Clustering of Applications with Noise (DBSCAN) \citep{dbscan} which works by identifying dense regions in the space by checking for the existence of a desired number of minimum objects (\textit{min\_samples}) within a desired distance (\textit{eps}) from a core point, which indicates a dense region. If they are found, it expands the cluster and repeats the same process until it cannot expand the dense region. Each dense region becomes a cluster, hence, the number of clusters is not known a priori. Points outside the dense regions do not belong to any cluster. For the latter, we employ agglomerative hierarchical clustering with complete linkage \citep{hclust}. It works by treating each point as a cluster of its own and iteratively merging closest clusters until it ends up at a single cluster. The distance between two clusters is the distance between their furthest-apart members. The whole process reveals a full hierarchical structure. A desired number of clusters can be extracted from this hierarchy at corresponding cut levels. Both methods are previously employed for embeddings that have angular separation \citep{bruno2019community} such as ours since our loss function $\mathcal{L}$ is based on $cossim()$ function which provides angular separations.

Both methods require the distance between two points but do not require their individual coordinates. We calculate the distance between two cities as $1-cossim(z_i,z_j)$ since cosine similarity is the function that gives the similarity in the latent space. In this way, we obtain distances between all city pairs within the range $[0,2]$. Using the calculated distance values as input, we perform our clustering analysis. For DBSCAN, we use hyperparameter values $0.14$ for \textit{eps} and $3$ for \textit{min\_samples} as initial experiments showed that this configuration (and also other configurations with nearby values) provides plausible clustering results.

\section{Results and findings}\label{sec:res}

\begin{figure}[!h]
  \centering
  \begin{subfigure}[b]{0.45\textwidth}
   \includegraphics[scale=0.36]{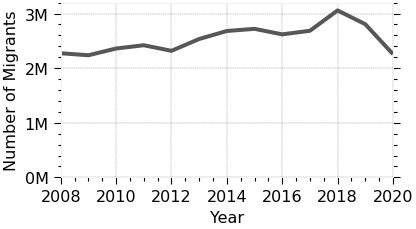}
  \caption{Migration volume}
  \label{subfig:migvoltrend}
  \end{subfigure}
  \begin{subfigure}[b]{0.45\textwidth}
   \includegraphics[scale=0.36]{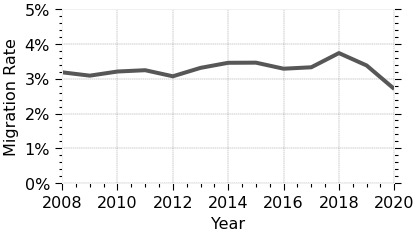}
  \caption{Migration rate}
  \label{subfig:migratetrend}
  \end{subfigure}
  \caption{Internal migration trend in Turkey}
  \label{fig:migtrend}
\end{figure}

The number of migrants over the years are presented in Figure \ref{fig:migtrend}. Each year, 2.2 million to 3 million migrate internally within Turkey, a country whose population increased from 71.5 million in 2008 to 83.6 million in 2020 \citep{tuik2021}. In terms of the total number of migrants, we observe a slow but often steady increase. However, this increasing trend is associated with the increasing population as the migration rate is mostly steady. Each year, 3\% to 3.5\% of the population migrate. The sudden decrease after 2019 is likely due to the COVID-19 pandemic as it limited mobility, e.g., almost all universities switched to distant learning, therefore many freshmen students have not relocated. Overall, while the results presented in the figure show an increase in volume, the migration rate does not show a notable increase. Therefore, these results support neither the law stating that migration rate increases nor the neoclassical argument suggesting that migration causes convergence of economic differences and hence the migration itself diminishes over time.

\begin{figure}[!h]
  \centering
  \begin{subfigure}[b]{0.3\textwidth}
   \includegraphics[scale=0.3]{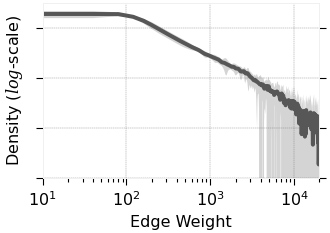}
  \caption{Distribution of edge weights}
  \label{subfig:weightdist}
  \end{subfigure}
  \begin{subfigure}[b]{0.3\textwidth}
   \includegraphics[scale=0.3]{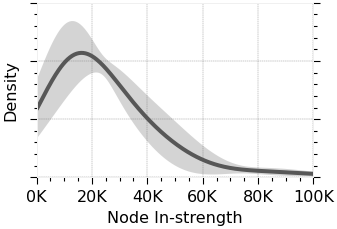}
  \caption{Distribution of node in-strengths}
  \label{subfig:instrdist}
  \end{subfigure}
  \begin{subfigure}[b]{0.3\textwidth}
   \includegraphics[scale=0.3]{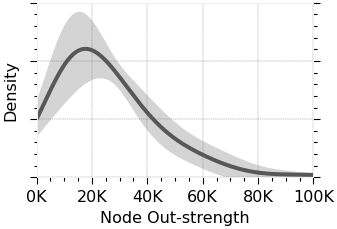}
  \caption{Distribution of node out-strengths}
  \label{subfig:outstrdist}
  \end{subfigure}
  \caption{Migration flow volume characteristics}
  \label{fig:multiscale}
\end{figure}

Figure \ref{fig:multiscale} shows the distribution of edge weights, node in-strenghts, and node out-strenghts of the original network. The solid lines denote the mean values over 13 years and the transparently colored regions around them indicate the area within the $2.58$ standard deviation.  Figure \ref{subfig:weightdist} reveals that edge weights follow a power law in the tail where a much smaller number of links have much larger weights. Figure \ref{subfig:instrdist} and \ref{subfig:outstrdist} reveal that node strenghts are also very heterogeneous where a smaller number of cities receive and send larger migration. We also note that the maximum values on the x-axes originally reach to $24884$, $328632$, and $381654$ respectively in Figure \ref{subfig:weightdist}, \ref{subfig:instrdist}, and \ref{subfig:outstrdist} respectively but cut at lower values for visualization purposes. In other words, the tails of the distributions are actually longer than they appear in the figures. These findings are also in line with the observation that most empirical networks exhibit a multiscale nature in their edge weights and node strengths. Moreover, it confirms the migration law that suggests sending and receiving locations in migration systems tend to follow a power law in their migration flow sizes.

For internal migration networks of Turkey, density values range from $0.999$ to $1$ over the years, which indicates that there are migration flows between all pairs of cities in both directions barring rare exceptions. Such networks are not suitable for most network analysis methods as many methods desire or require sparsity. Hence, the need for filtering out the insignificant links is clear. Following the need for information filtering methods to extract the sparse backbone of the network while respecting the multiscale nature of the network, we extract the signed backbone of networks using significance filter and vigor filter which are shown to extract statistically sound and meaningful backbones of intrinsically dense networks

\begin{figure}[!h]
\centering
  \begin{subfigure}[b]{0.45\textwidth}
   \includegraphics[scale=0.36]{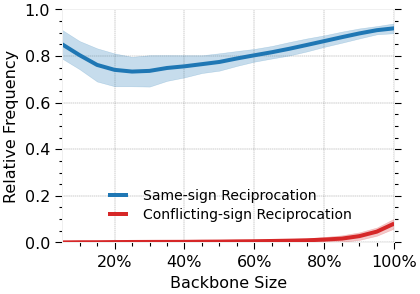}
  \caption{Reciprocity}
  \label{subfig:rec}
  \end{subfigure}
  \begin{subfigure}[b]{0.45\textwidth}
   \includegraphics[scale=0.36]{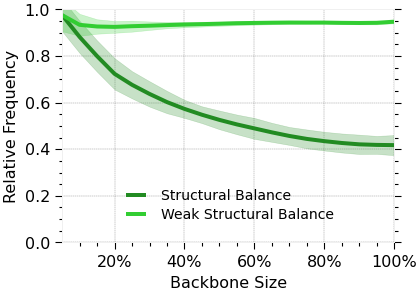}
  \caption{Structural balance}
  \label{subfig:sb}
  \end{subfigure}
  \caption{Characteristics of migration backbones}
  \label{fig:sbrec}
\end{figure}

We extract the signed backbones via the significance filter retaining 5\% to 100\% of all possible links and evaluate the structure of backbones in terms of reciprocity and structural balance.  Figure \ref{subfig:rec} visualizes the proportion of positive-positive and negative-negative reciprocated links as well as the proportion of conflicting links over all links. It shows that most signed links are reciprocated with links of the same sign. Conflicting links are rare even when the backbone size is considerably large. These results indicate that the link signs in our migration networks are consistent in terms of reciprocity. This is further in coherence with the law on counter-streams which states that for every major migration stream, there is a counter-stream. In a sense, our backbone extraction method specifically retains the major streams only, hence the analysis on the extracted backbone is particularly appropriate for this test. Going beyond the explicit statement of this law, it also shows that for major negative streams (i.e., negative links where migrants from a particular city avoid moving to a particular city), there are negative streams in the opposite direction. Also, the lack of conflicting links shows that for every major stream, no negative streams exist in the opposite direction; if otherwise, it would invalidate the law.

Next, we investigate the structural balance and weak structural balance of the extracted backbones. As these measures are defined for undirected networks, we convert our directed migration networks to undirected versions by assuming a positive link when all directed links between a pair are positive, a negative when all directed links between a pair are negative; and assuming no link if there is one positive and one negative directed link between a pair. When there is only a single link between a pair, we assume an undirected link of the same sign. Figure \ref{subfig:sb} reports the SB and WSB results. It demonstrates great levels of SB when the backbone is sufficiently sparse and great levels of WSB even if the network very dense. These results indicate that the nodes in our networks have structural balance. Such structural balance may arise due to positive relationships among geographically close cities as well as their common destinations in longer distances, e.g., three neighboring cities send migration to each other and they also send migration to a specific industrial city in long distance. Visualizations that are provided later in this study provide support for this statement.

Overall, Figure \ref{fig:sbrec} shows that extracted backbones have characteristics usually observed in other empirical signed networks; the findings are intuitive and are coherent with the relevant findings in the migration network literature. This conclusion also justifies our choice in selecting an appropriate information filtering (i.e., backbone extraction) method.

\begin{figure}[!h]
    \centering
    \begin{subfigure}[b]{0.9\textwidth}
    \includegraphics[width=\textwidth]{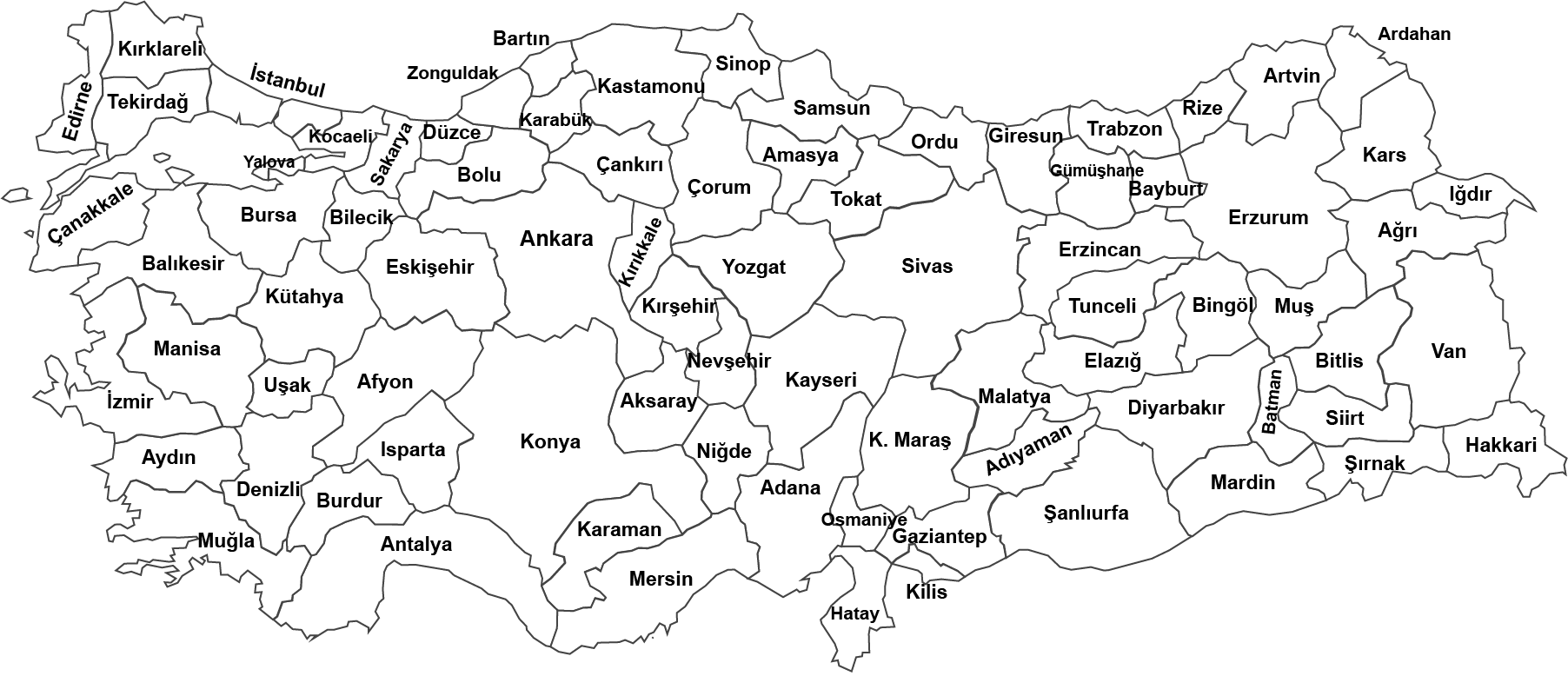}
    \caption{Map of cities in Turkey}
    \label{subfig:netmig}
    \end{subfigure}
    \vspace{10pt}
    
    \begin{subfigure}[b]{0.45\textwidth}
    \includegraphics[width=\textwidth]{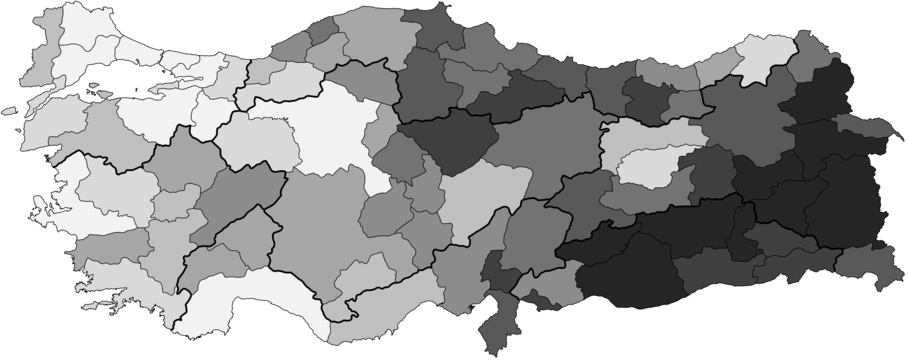}
    \caption{GDP per capita}
    \label{subfig:gdppercapita}
    \end{subfigure}
    \begin{subfigure}[b]{0.45\textwidth}
    \includegraphics[width=\textwidth]{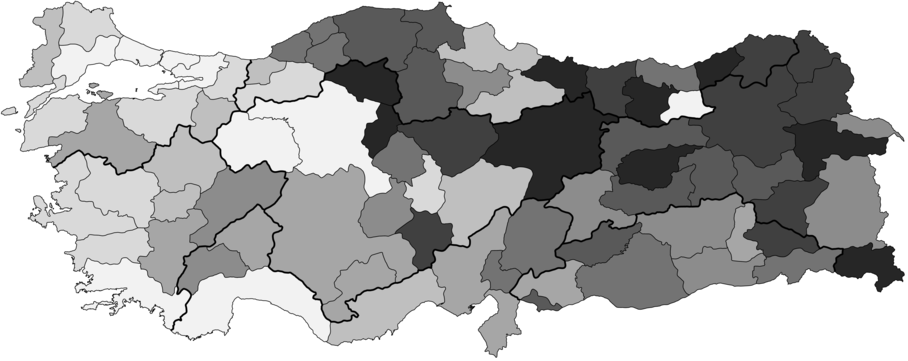}
    \caption{Net migration rate}
    \label{subfig:netmigration}
    \end{subfigure}

  \caption{Maps of Turkey}
  \label{fig:infomap}
\end{figure}

Figure \ref{fig:infomap} presents the map of Turkey with city borders as well as distribution of GDP per capita and net migration rate in 2019. Observing this information will also help readers that are not familiar with socioeconomic differences in Turkey by providing a context. In Figure \ref{subfig:gdppercapita}, the increasing values are shown with gray colors transitioning from dark to light with each color denoting an equal-size group. In a similar fashion, in Figure \ref{subfig:netmig}, those cities with increasing net migration rate (i.e., in-migrants minus out-migrants divided by the city population) are denoted with colors transitioning from dark gray to light gray. Overall, the figure demonstrates that GDP per capita increases in general as we move towards the west and net migration rate increases as we move toward the east. This supports the view that migration is tied to economical differences between locations.

For the rest of the analysis, before we proceed with the representation learning methods, we extract the backbones for each year in the following way. We first retain 7.5\% of links that are most significant and further eliminate links that have absolute vigor values less than $0.33$. Vigor threshold is selected as such so that the weight of a positive link is at least more than double the random expectation and the weight of a negative link is at most the half of random expectation. Therefore, the backbones contain the significant and intense links.

\begin{figure}[!h]
  \centering
  \begin{subfigure}[b]{0.9\textwidth}
  \includegraphics[width=\textwidth]{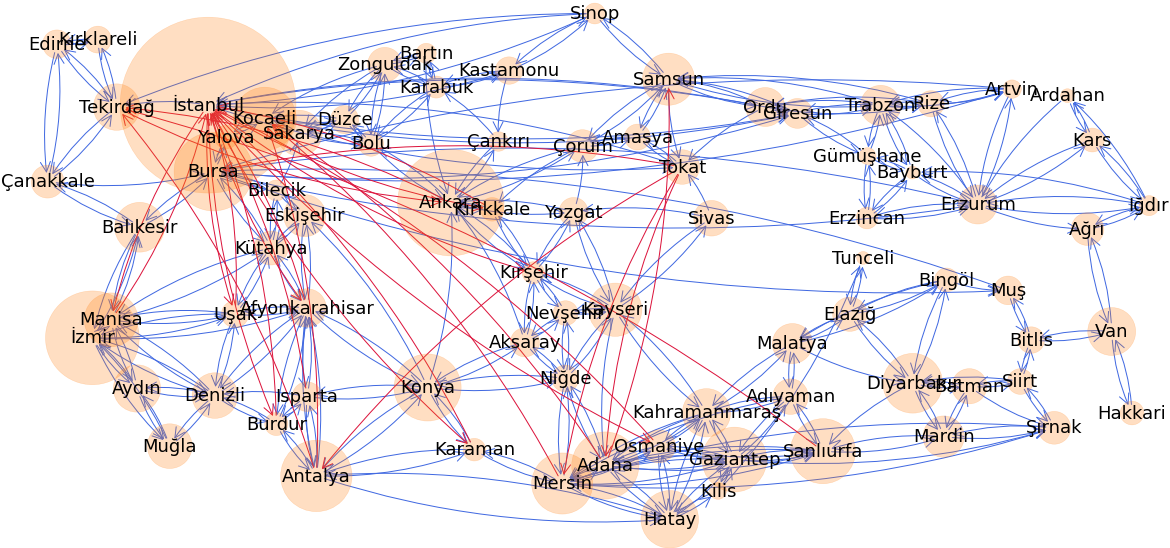}
  \caption{Overall network}
  \label{subfig:overall_2020}
  \end{subfigure}
  \vspace{10pt}

  \begin{subfigure}[b]{0.3\textwidth}
   \includegraphics[width=\textwidth]{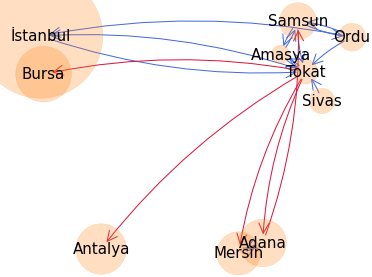}
  \caption{Tokat ego network}
    \label{subfig:tokat_2020}
  \end{subfigure}
    \begin{subfigure}[b]{0.3\textwidth}
    \includegraphics[width=\textwidth]{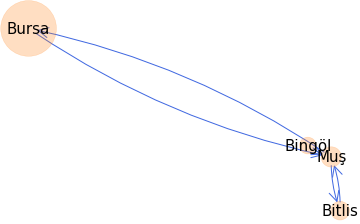}
  \caption{Muş ego network}
  \label{subfig:mus_2020}
  \end{subfigure}
    \begin{subfigure}[b]{0.3\textwidth}
    \includegraphics[width=\textwidth]{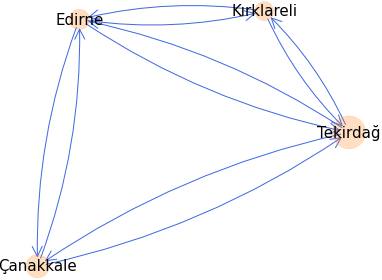}
  \caption{Edirne ego network}
  \label{subfig:edirne_2020}
  \end{subfigure}
  \caption{Spatial network of migration backbone in 2020}
  \label{fig:backbone2020}
\end{figure}

\begin{figure}[!h]
  \centering
  \begin{subfigure}[b]{0.9\textwidth}
   \includegraphics[width=\textwidth]{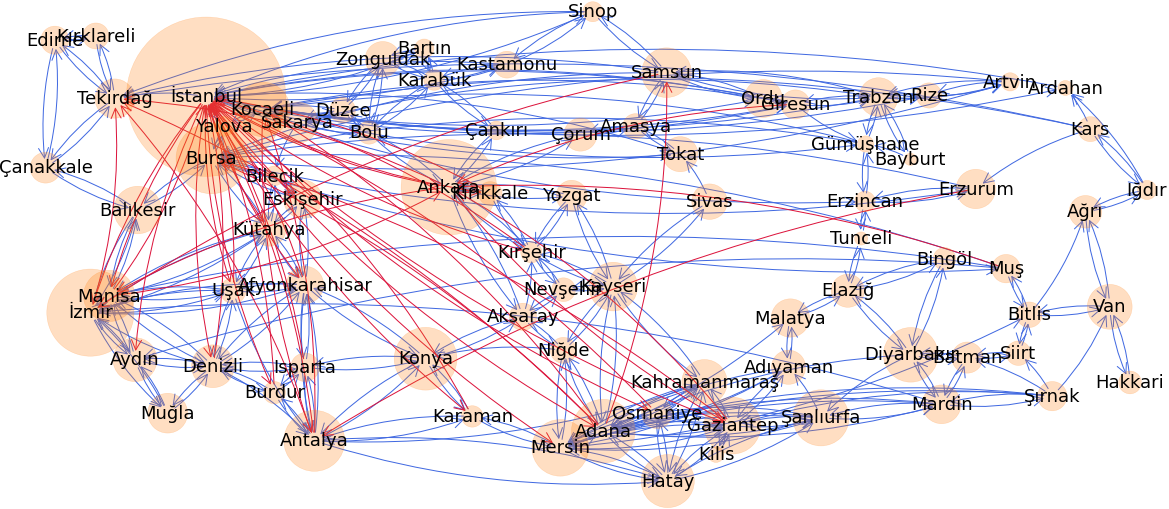}
  \caption{Overall network}
  \label{subfig:overall_2008}
  \end{subfigure}
  \vspace{5pt}

  \begin{subfigure}[b]{0.3\textwidth}
   \includegraphics[width=\textwidth]{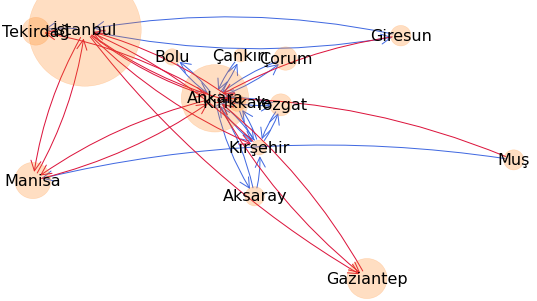}
  \caption{Ankara ego network}
    \label{subfig:ankara_2008}
  \end{subfigure}
    \begin{subfigure}[b]{0.3\textwidth}
   \includegraphics[width=\textwidth]{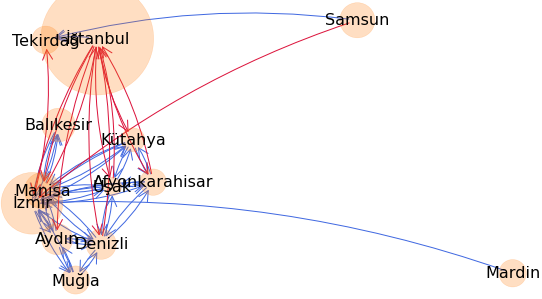}
  \caption{Izmir ego network}
  \label{subfig:izmir_2008}
  \end{subfigure}
    \begin{subfigure}[b]{0.3\textwidth}
   \includegraphics[width=\textwidth]{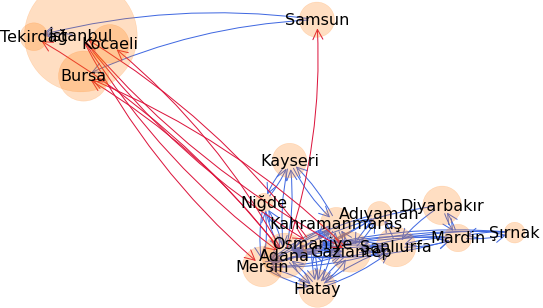}
  \caption{Adana ego network}
  \label{subfig:adana_2008}
  \end{subfigure}
  \caption{Spatial network of migration backbone in 2008}
  \label{fig:backbone2008}
\end{figure}

Figure \ref{subfig:overall_2020} and \ref{subfig:overall_2008} visualizes the extracted backbones in a spatial geographic layout, respectively for 2008 and 2020. The node sizes are proportional to their population in respective years. In both figures, most positive links are local within close spatial proximity. Most of these local links are reciprocated. There are no local negative links. The non-local links that connect distant locations are relatively few and usually towards cities with great economic and social activity such as Istanbul. Most of the less local positive links are in the east-west direction. This may be explained by the fact that historically large and economically more active cities are concentrated in the western parts. Also, the mountain ranges in northern and southern regions in Turkey lay in the east-west direction which means that historically, most communication and transportation are developed in a horizontal direction. In addition to the positive examples, negative examples also support this finding as the negative links are usually and loosely in the north-south direction. These overall results support some of the previously described rules, e.g., migration takes place in well-defined routes, most migration happens at a local level, and long-range migrations usually involve centers of large economic activity.

We also investigate ego networks for selected cities for more particular analysis. For 2020, we present the ego networks for Tokat, Muş, and Edirne respectively in Figure \ref{subfig:tokat_2020}, \ref{subfig:mus_2020}, and \ref{subfig:edirne_2020} by visualizing both ego-alter links as well as alter-alter links. Tokat has positive links with most of its geographically close neighbors in addition to a positive link with Istanbul, a distant city with a traditionally large socio-economical activity. In contrast, it has negative links to few southern cities that otherwise attract relatively large emigration. Muş shows a similar behavior with positive connections to its geographically closer neighbors in addition to a positive link with a distant but economically more active larger city, Bursa. In contrast to these two cities with low economic activity, Edirne has a relatively good economic situation. Accordingly, Edirne does not have long-range links and all of its links are positive and connected to its immediate geographical neighbors. The specific analyses on the three cities and the general picture provided earlier indicate that certain smaller or economically less active cities have migration outflows towards different and specific large and economically active cities rather than homogeneous outflows.

Such preference of Tokat towards Istanbul and Muş towards Bursa may be explained by the previous strong social ties between the respective pairs. A proxy for such social ties is the number of people who live in the city with greater economic activity but who were born in the city with lesser economic activity. In 2020, based on the data from \cite{tuik2021}, among those who were born in Tokat but live elsewhere, 52\% live in Istanbul followed by Ankara with only 8\%. The same figure for Muş and Bursa is 16\% with Bursa being the second most popular destination after Istanbul. Among those who live in Istanbul but were born in other cities of Turkey, 4\% is born in Tokat with Tokat being the second most popular origin. The same figure for Bursa and Muş is 5\% with Muş being the 4th most popular origin. Therefore, the claim about the effects of existing social ties is supported. Furthermore, in both city pairs, the links are reciprocated which supports the migration law that counter-streams develop for every major stream.

In Figure \ref{subfig:ankara_2008}, \ref{subfig:izmir_2008}, and \ref{subfig:adana_2008}, for 2008, we visualize the ego networks of second, third, and fifth-most populous cities in Turkey: Ankara, İzmir, and Adana.  We observe that Ankara has its positive links exclusively with relatively nearby cities. As the capital city of Turkey, it seems to attract migration relatively homogeneously, a deviation from the behavior of other large cities which is worthy of further investigation. It also has negative links with specific smaller cities, such as Muş and Giresun, which have well-defined streams to other large cities. Other negative links are with relatively larger-economy cities such as Istanbul, Manisa, and Gaziantep. This indicates that number of migrants between populous and economically-well cities are lower than randomly expected, which further hints that the migrants seek to increase their economic utility. The ego network of Izmir indicates mostly local positive links with very high mobility among the cities in the region. The only long-range positive link is from Mardin. Following the social ties hypothesis, we find that among those who were born in Mardin but live elsewhere, 16\% live in Izmir with Izmir being the second most popular destination after Istanbul. Among those who live in Izmir but were born in another city, Mardin is the third most popular origin with 4\%. Ego network for Adana shows that it plays the role of a regional hub, as is the case with Izmir. It has negative links to the northwestern region, another and larger hub which attracts many long-range links from northern parts of the country.

Due to space limitations, we cannot visually investigate overall networks or ego networks for all years or all cities. Our unreported investigations and the stability results presented earlier indicate that major findings stay the same over the years. Interested readers may conduct their own analyses easily for other years following the same techniques\footnote{The code for reproducing the results will be published once and if the manuscript passes the double-blind peer review process and is accepted for publication.}.

\begin{figure}[!h]
    \centering
    \begin{subfigure}[b]{0.45\textwidth}
    \includegraphics[width=0.9\textwidth]{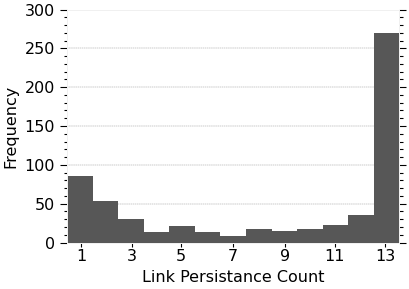}
    \caption{}
    \label{subfig:lp_overall}
    \end{subfigure}
    \begin{subfigure}[b]{0.45\textwidth}
    \includegraphics[width=0.9\textwidth]{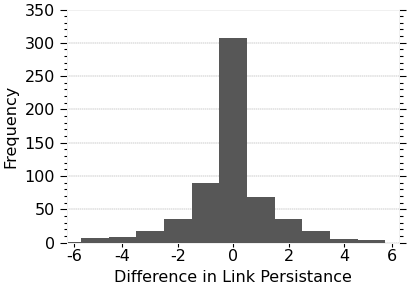}
    \caption{}
    \label{subfig:lp_temporal}
    \end{subfigure}
  \caption{Link persistence in migration backbones}
  \label{fig:linkpersistance}
\end{figure}

Following the observation that the overall migration backbone has not notably changed from 2008 to 2020, we also look at the persistence of individual links over time. Over 13 time steps, we look at how many times each directed link is repeated. 606 links appear at least once. For those links appearing at least once, Figure \ref{subfig:lp_overall} shows that most of the links persist in all years with a much smaller but notable portion of links appearing only once or twice. We also compare the number of links during two equal periods, 2008-2013 and 2014-2019. We exclude 2020 since it shows a distinct behavior, most likely due to the COVID-19 pandemic. 597 links appear at least once in at least one period. Then, we calculate the difference between frequencies observed in two time periods for those 597 links. Figure \ref{subfig:lp_temporal} shows that most of the links have equal frequencies in both years where a very small number of cities appear frequently in one period and appear only a few times (or does not appear at all) in the other period.

\begin{figure}[!h]
\centering
\includegraphics[scale=0.35]{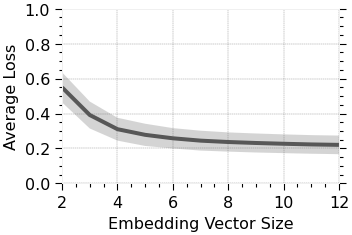}
\caption{Embedding loss}
\label{fig:embloss}
\end{figure}

Next, we move to our analysis of the underlying latent low-dimensional space. We generate embeddings for each $d$ in $\{2, 3, 4, \dots, 12\}$. Figure \ref{fig:embloss} shows the average embedding loss over all links, i.e., all nondiagonal elements of $W$. The solid line denotes the mean value over 13 years and the transparently colored region indicates the area within its $2.58$ standard deviation. As $d$ increases, the marginal gain in preserved information quickly disappears. We choose $d=8$ for the rest of the analysis.

\begin{figure}[!h]
\centering
\includegraphics[width=0.55\textwidth]{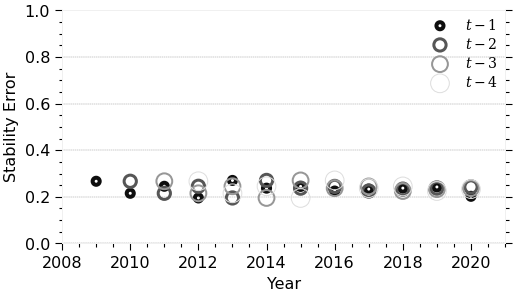}
\caption{Stability error over time}
\label{fig:staberr}
\end{figure}

To test our earlier findings on the dynamics of the migration system, i.e., the general stability observed in the overall structure of the network and most of the links, we investigate the overall changes in the embedding space over time using stability error.  Figure \ref{fig:staberr} presents the stability error over time by comparing each $Z^t$ with $Z^{t-1}$, $Z^{t-2}$, $Z^{t-3}$, and  $Z^{t-4}$. The results demonstrate that the stability error is generally low and does not fluctuate considerably over time - even when compared to the system from four years earlier. It can be concluded that the structure of the migration network is largely stable over the investigated period and its overall dynamics seem to be modest without any critical transition points. This finding is also in line with the migration law stating that migration streams are well-defined.

\begin{figure}[!h]
    \centering
    \begin{subfigure}[b]{0.32\textwidth}
    \includegraphics[width=\textwidth]{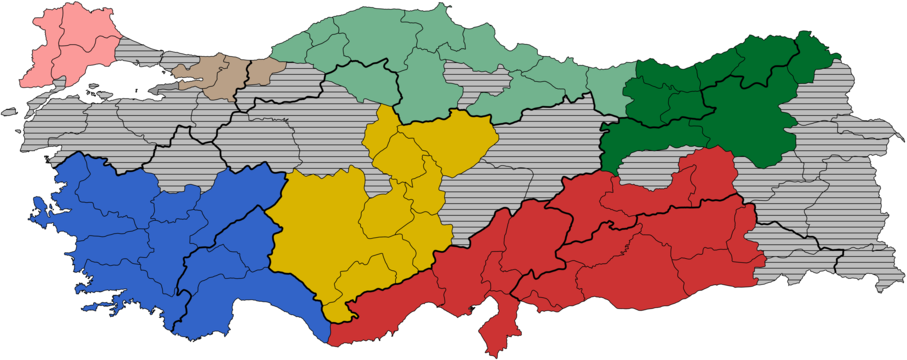}
    \caption{2008}
    \end{subfigure}
    \begin{subfigure}[b]{0.32\textwidth}
    \includegraphics[width=\textwidth]{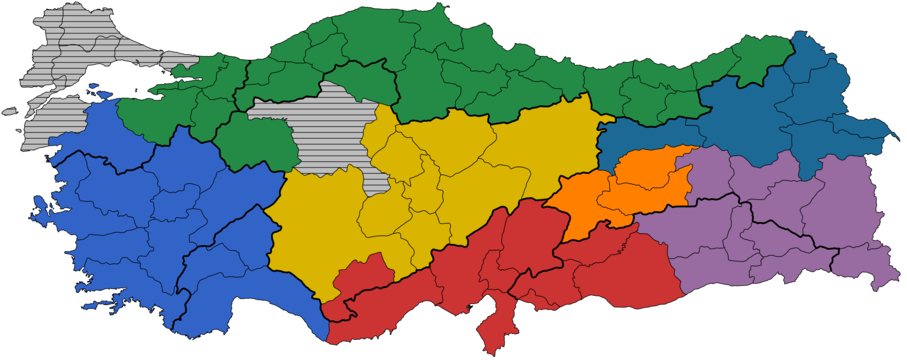}
    \caption{2014}
    \end{subfigure}
    \begin{subfigure}[b]{0.32\textwidth}
    \includegraphics[width=\textwidth]{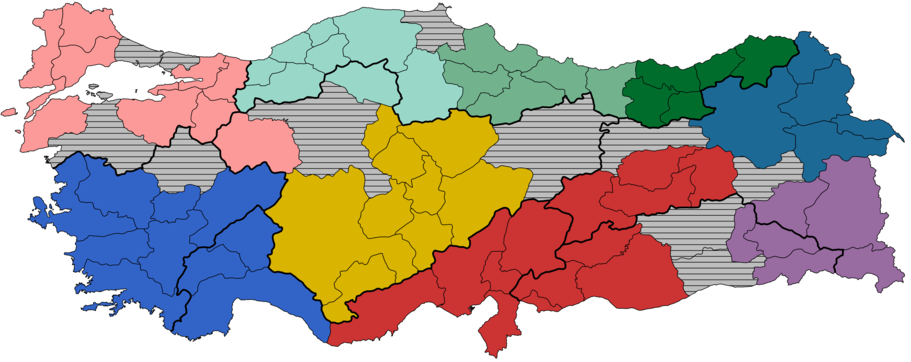}
    \caption{2020}
    \end{subfigure}
  \caption{Density-based clusters}
  \label{fig:latentclusters}
\end{figure}

Figure \ref{fig:latentclusters} visualizes the clustering results in 2008, 2014, and 2020 using different colors for each cluster. Cities that do not belong to any cluster are colored in gray with a horizontal grid pattern. In line with previous results, it shows that geographical proximity plays an important role. However, such geographical proximity is not strictly bounded by the 7 geographical regions whose borders are separated with thicker lines. Density-based clustering results demonstrate that dense clusters are in line with geographical proximity. However, some cities do not belong to any cluster, which indicates that the positions of these cities are not bounded by their geographic positions. The most prominent example is Istanbul that has many long-range links to different geographies. Another example is Ankara, the capital city. We also note that the found communities and the particular cases of Istanbul and Ankara were recognized in the findings of \cite{slater1975identification}, which further suggests that the migration dynamics have not significantly changed in the last several decades. We also observe that cluster boundaries are fluid with some cities in their borders belonging to different clusters across time. This may be explained not only by their dynamically changing behavior but also by the fact that they are not strongly attached to their corresponding clusters. On the other hand, particularly the cluster in the west denoted with blue color, have its most members same across time. We should also note that it is possible to obtain different but congruent communities, i.e., more or less granular communities, using different hyperparameters for the clustering method.

\begin{figure}[!h]
  \centering
  \begin{subfigure}[b]{0.31\textwidth}
   \includegraphics[width=\textwidth]{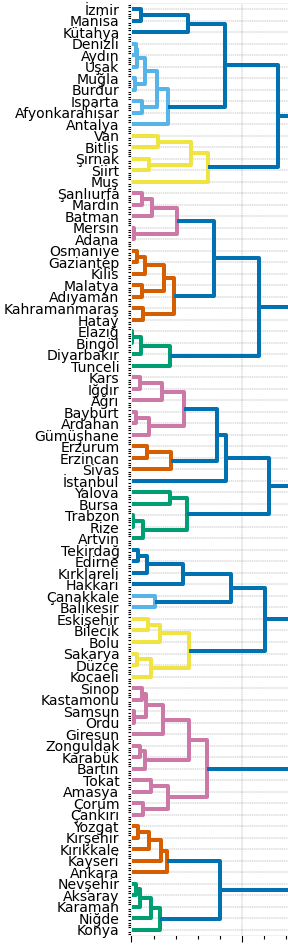}
  \caption{2008}
  \end{subfigure}
\begin{subfigure}[b]{0.31\textwidth}
   \includegraphics[width=\textwidth]{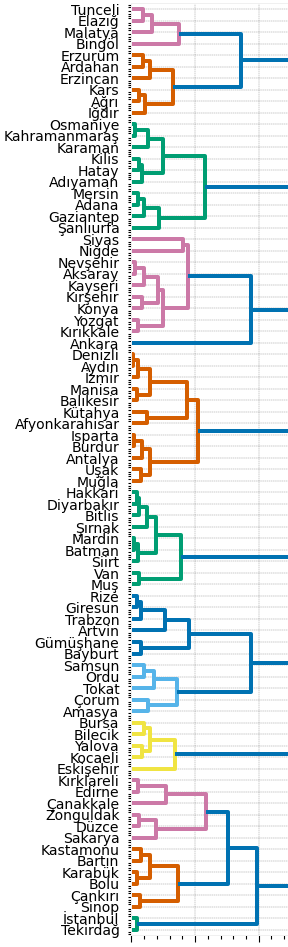}
  \caption{2014}
  \end{subfigure}
  \begin{subfigure}[b]{0.31\textwidth}
   \includegraphics[width=\textwidth]{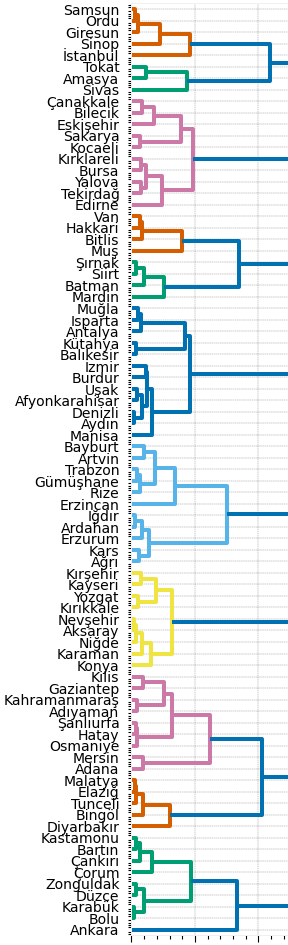}
  \caption{2020}
  \end{subfigure}
  \caption{Hierarchical clustering dendrograms}
  \label{fig:hclust}
\end{figure}

We also investigate hierarchical clusters in Figure \ref{fig:hclust} for 2008, 2014, and 2020. As we employ the complete-linkage method which requires most distant members to be sufficiently similar, we do not produce the full dendrogram but only present the smaller communities at the bottom of the hierarchy. The figures, in all years, demonstrate that geographical proximity plays a major role in defining communities. However, Istanbul, which does not belong to any cluster in the density-based result and has many long-range links, is often clustered together with cities that it does not share borders with. Overall, the results from both clustering methods tend to agree, which shows that the findings from the clustering methods are robust.

\section{Conclusion}\label{sec:con}
In this study, we have investigated the internal migration patterns of Turkey from 2008 to 2020 via signed network analysis, ego network analysis, representation learning, temporal stability analysis, community detection, and network visualization. We find that most migration links are geographically bounded with exceptional long-range links involving centers of large economic activity. Major migration flows are reciprocated with migration flows in the opposite direction. The migration flows take place on well-defined routes and the migration system is largely stable over the investigated time period. Overall, our analysis and presented findings indicate that the internal migration system of Turkey agrees with the general migration laws and provide unique insights into the internal migration patterns of Turkey.

\backmatter

\bibliography{sn-bibliography}

\end{document}